\documentclass[graybox]{svmult}

\usepackage{mathptmx}       
\usepackage{helvet}         
\usepackage{courier}        
\usepackage{type1cm}        
\usepackage{makeidx}         
\usepackage{graphicx}        
\usepackage{multicol}        
\usepackage[bottom]{footmisc}

\makeindex             

\newcommand{\be}{\begin{equation}}
\newcommand{\ee}{\end{equation}}
\newcommand{\bea}{\begin{eqnarray}}
\newcommand{\eea}{\end{eqnarray}}
\newcommand{\nn}{\nonumber}

\font\tenscr=rsfs10 scaled1100
\font\sevenscr=rsfs7 
\font\fivescr=rsfs5 
\skewchar\tenscr='177
\skewchar\sevenscr='177
\skewchar\fivescr='177
\newfam\scrfam
\textfont\scrfam=\tenscr
\scriptfont\scrfam=\sevenscr
\scriptscriptfont\scrfam=\fivescr

\begin{document}

\title*{Area inequalities for stable marginally trapped
surfaces}
\author{Jos\'e Luis Jaramillo}
\institute{Jos\'e Luis Jaramillo \at Max-Planck-Institut f{\"u}r Gravitationsphysik, Albert Einstein
Institut, Am M\"uhlenberg 1 D-14476 Potsdam Germany, \email{Jose-Luis.Jaramillo@aei.mpg.de}}
%
%
\maketitle

\abstract*{}

\abstract{
We discuss a family of inequalities involving the area, angular momentum
and charges of stably  outermost marginally trapped surfaces in generic non-vacuum
dynamical spacetimes, with non-negative cosmological constant and  matter sources
satisfying the dominant energy condition.
These inequalities provide lower bounds for the area 
of spatial sections of dynamical trapping horizons, namely hypersurfaces 
offering quasi-local models of black hole horizons.
In particular, these inequalities represent particular examples of the extension
to a Lorentzian setting of tools employed in the discussion of 
minimal surfaces in Riemannian contexts.
}

\section{Introduction}
The Lorentzian nature of spacetime geometry, with its inherent notion of
null cone, controls the rich features of light bending in general relativity.
This includes in particular the possibility of causal disconnection between spacetime 
regions, as well as the convergence behavior of (trapped)
light rays. Both aspects, related by the notion of (weak) cosmic censorship
\cite{Pen69}, 
lay at the basis of the concept of black hole in general relativity.
In spite of the complexity of the generic situation, it is
remarkable that stationary and vacuum black hole spacetimes are completely characterized 
by a few parameters with physical interpretation, namely the mass $M$ (or, alternatively, 
the area $A$ of the horizon), the angular momentum $J$ and certain charges $Q_i$. 
These parameters fulfill a class of geometric inequalities that bound the mass by below. 
When using the horizon area $A$ instead of the mass, they present the general form
\bea
\label{e:inequality}
(A/(4\pi))^2\geq (2J)^2 + (\sum_i Q_i^2)^2 \ ,
\eea
and provide a family of inequalities written completely in terms 
of the quasi-local geometry of the black hole horizon.
As a second remarkable point, these quasi-local geometric inequalities  
do extend to fully generic dynamical and non-vacuum situations, providing 
general lower bounds for the black hole horizon area.

In the stationary axisymmetric case with matter surrounding the horizon, 
these quasi-local inequalities have been proved 
to hold for the Killing horizon in \cite{AnsPfi07,HenAnsCed08,Hennig:2008zy,Ansorg:2010ru}.
Regarding dynamical situations, and after the study in 
\cite{Dai06a,Dain:2005qj,Dain:2006wb,Chrus07a,Chrus07b,Costa:2009hn,Chrusciel:2009ki}
of the global vacuum axisymmetric inequalities involving $M$, the quasi-local 
vacuum case has been studied in
\cite{Dain:2010qr,Acena:2010ws,GabachClement:2011kz,Dain:2011pi} using axisymmetric
initial data.
Finally, in  \cite{Jaramillo:2011pg,Dain:2011kb,Clement:2011tq,GabJarRei11} a purely 
spacetime (Lorentzian) perspective has been adopted, permitting to identify and refine the
key assumptions, this leading to the extension of the inequalities to generic dynamical
scenarios with matter.
The study of these inequalities in higher dimensions has been started in
\cite{Hollands:2011sy}. A general review on geometric 
inequalities in axially symmetric black holes is presented in \cite{Dain:2011mv} .

Here we discuss these quasi-local inequalities, placing the emphasis on the involved 
Lorentzian aspects, namely the notion of stability of marginally outer trapped surfaces.
The latter provide a Lorentzian analogue to the notion
of stable minimal surfaces in Riemannian geometry. 
New results are presented regarding the incorporation in the inequalities of 
Yang-Mills charges and a 
geometric charge for certain divergence-free vectors on closed surfaces. We also
comment on the interpretation
of the integral characterization of the stability condition as an {\em energy flux} inequality.

\section{Geometric and physical elements}
\label{s:geomelem}
Let $(M, g_{ab})$ be a $4$-dimensional Lorentzian manifold
satisfying Einstein equations
\bea
\label{e:Einstein_eq}
G_{ab} + \Lambda g_{ab} = 8\pi T_{ab} \ ,
\eea
where $G_{ab} = R_{ab} -\frac{1}{2}Rg_{ab}$ is the Einstein tensor
associated with the Levi-Civita connection $\nabla_a$,
$\Lambda$ is the cosmological constant and $T_{ab}$ the stress-energy
tensor. Unless otherwise stated, in the following the stress-energy tensor 
is assumed to satisfy a dominant
energy condition (namely, given a future-directed causal vector $v^a$, 
then $-{T^a}_b v^b$ is a future-oriented causal vector)
and the cosmological constant to be non-negative
$\Lambda\geq 0$. We use throughout
Penrose's abstract index notation (e.g. \cite{Wal84}).

\subsection{Geometry of 2-surfaces}
\label{s:2surfaces}
Let us consider a closed orientable 2-surface ${\cal S}$ embedded in $(M, g_{ab})$
(in the following, we shall assume that surfaces ${\cal S}$ are closed and orientable,
unless otherwise stated).
Let us denote the induced metric on ${\cal S}$ as $q_{ab}$, with Levi-Civita
connection $D_a$, Ricci scalar ${}^2\!R$ and volume element $\epsilon_{ab}$ 
(we will denote by $dS$ the area measure on ${\cal S}$). 
Decomposing the tangent plane $T_p M$ at each point $p\in {\cal S}$ as
$T_p M = T_p{\cal S}\oplus T_p^\perp{\cal S}$,
let us consider null vectors $\ell^a$ and $k^a$ spanning the normal plane $T_p^\perp{\cal S}$ and
normalized as $\ell^a k_a = -1$. This leaves a (boost) rescaling freedom 
$\ell'^a =f \ell^a$, $k'^a = f^{-1} k^a$. We can write 
\bea
\label{e:metric_q}
q_{ab} = g_{ab} +k_a \ell_b + \ell_a k_b \ .
\eea
Regarding the extrinsic curvature elements that we need in our analysis, let us consider 
the deformation tensors $\Theta^{(\ell)}_{ab}$ and $\Theta^{(k)}_{ab}$ along  
$\ell^a$ and $k^a$, respectively
\bea
\Theta_{ab}^{(\ell)}\equiv{q^c}_a {q^d}_b \nabla_c \ell_d \ \ , \ \ 
\Theta_{ab}^{(k)}\equiv{q^c}_a {q^d}_b \nabla_c k_d \ .
\eea
They determine the second fundamental form ${\cal K}_{ab}^c$ of $({\cal S}, q_{ab})$ into
$(M, g_{ab})$, namely ${\cal K}_{ab}^c\equiv {q^d}_a {q^e}_b \nabla_d {q^c}_e = 
k^c\Theta_{ab}^{(\ell)} + \ell^c\Theta_{ab}^{(k)}$ (cf. Senovilla's contribution in this volume).
In particular, the expansion $\theta^{(\ell)}$ and the shear  $\sigma^{(\ell)}_{ab}$ 
associated with the null normal $\ell^a$, are given respectively by the trace and traceless 
parts of $\Theta_{ab}^{(\ell)}$
\bea
\label{e:expansion_shear}
\theta^{(\ell)}\equiv q^{ab} \Theta_{ab}^{(\ell)}= q^{ab}\nabla_a\ell_b \ \ , \ \  
\sigma^{(\ell)}_{ab} \equiv  \Theta_{ab}^{(\ell)} - \frac{1}{2}\theta^{(\ell)}q_{ab} \ .
\eea
In addition, we consider the normal fundamental form $\Omega_a^{(\ell)}$ 
\bea
\label{e:Omega}
\Omega^{(\ell)}_a = -k^c {q^d}_a \nabla_d \ell_c \ ,
\eea
that provides a connection on the normal bundle $T^*_\perp {\cal S}$. More specifically,
considering a form $v_a\in T^*_\perp {\cal S}$, expressed as $v_a = \alpha \ell_a + \beta \ell_b$,
we can write ${q^c}_a\nabla_c v_b = \Theta_{ab}^{(v)} +  D_a^\perp v_b$, where
$\Theta_{ab}^{(v)}={q^c}_a {q^d}_b \nabla_c v_d = \alpha \Theta_{ab}^{(\ell)}  + \beta\Theta_{ab}^{(k)}$
and
\bea
\label{e:normalconnection1}
D_a^\perp v_b =  D_a^\perp (\alpha \ell_a + \beta \ell_b) = 
(D_a \alpha + \Omega^{(\ell)}_a \alpha )\ell_b + (D_a \beta + \Omega^{(\ell)}_a \beta )k_b \ .
\eea
Transformation rules under a null normal rescaling $\ell'^a =f \ell^a$, $k'^a = f^{-1} k^a$ are
\bea
\label{e:null_transformations}
\theta^{(\ell')}=f \theta^{(\ell)} \ \ , \ \ 
\sigma^{(\ell')}_{ab}= f\sigma^{(\ell)}_{ab} \ \ , \ \
\Omega^{(\ell')}_a = \Omega^{(\ell)}_a + D_a(\mathrm{ln}f).
\eea

\subsubsection{Axisymmetry}
\label{s:axisymmetry}
The introduction of a canonical angular momentum $J$ on ${\cal S}$ requires
imposing axisymmetry.
In this context, we require the geometry of ${\cal S}$ to be axisymmetric
with axial Killing vector $\eta^a$ on ${\cal S}$. 
More specifically we require
\bea
\label{e:axisymmetry}
{\cal L}_\eta q_{ab}=0 \ \ , \ \  {\cal L}_\eta \Omega_a^{(\ell)}=0 
\ \ , \ \ {\cal L}_\eta \ell^a = {\cal L}_\eta k^a = 0 \ ,
\eea
where $\eta^a$ has closed integral curves, vanishes exactly 
at two points on ${\cal S}$ and is normalized so that its integral curves 
have an affine length of $2\pi$. 
We adopt a tetrad  $(\xi^a, \eta^a, \ell^a, k^a)$ on ${\cal S}$,
where the unit vector $\xi^a$ tangent to ${\cal S}$ satisfies
$\xi^a\eta_a=\xi^a\ell_a=\xi^ak_a=0$, $\xi^a\xi_a=1$.
We can then write $q_{ab}=\frac{1}{\eta}\eta_a\eta_b + \xi_a\xi_b$, 
with $\eta=\eta^a\eta_a$, and 
\bea
\label{e:Omega_eta_xi}
\Omega^{(\ell)}_a = \Omega^{(\eta)}_a + \Omega^{(\xi)}_a \ \ \ \ , \ \ \ \ 
\Omega^{(\ell)}_a{\Omega^{(\ell)}}^a = \Omega^{(\eta)}_a{\Omega^{(\eta)}}^a + 
\Omega^{(\xi)}_a{\Omega^{(\xi)}}^a \ ,
\eea
with $\Omega^{(\eta)}_a= \eta^b\Omega^{(\ell)}_b \eta_a/\eta$
and $\Omega^{(\xi)}_a= \xi^b\Omega^{(\ell)}_b \xi_a$. 
We can introduce now a canonical (gravitational) angular momentum as
\bea
\label{e:Komar_angular_momentum}
J_{\mathrm K} = \frac{1}{8\pi}\int_{\cal S} \Omega_a^{(\ell)} \eta^a dS \ ,
\eea
where the divergence-free character of $\eta^a$ together with 
the transformations 
properties of $\Omega_a^{(\ell)}$ in (\ref{e:null_transformations})
guarantee the invariance of $J$ under a rescaling
of the null normals. 
This angular momentum on ${\cal S}$ coincides with the Komar one,
namely $J_{\mathrm{Komar}}=\frac{1}{8\pi} \int_{\cal S} \nabla_a \eta_b dS^{ab}$ with
$dS^{ab}=\frac{1}{2}(k^a\ell^b-\ell^a k^b)dS$, 
if $\eta^a$ can be extended as a  Killing vector to a spacetime
neighborhood of ${\cal S}$.

\subsection{Electromagnetic field}
Let us consider an electromagnetic field on $(M, g_{ab})$ with 
strength field  (Faraday) tensor $F_{ab}$. On a local chart 
we can express $F_{ab}$ in terms of a vector potential $A_a$ as 
$F_{ab}=\nabla_a A_b - \nabla_b A_a$. The electromagnetic
stress-energy tensor is given by
\bea
\label{e:EMstressenergy}
 T^{\mathrm{EM}}_{ab}= \frac{1}{4\pi}\left(F_{ac}
    F_b{}^{c}-\frac{1}{4}g_{ab} F_{cd} F^{cd}  \right) .
\eea
Given ${\cal S}$, we denote the electric 
and magnetic field components normal to ${\cal S}$ as
\bea
\label{e:E_B_normal}
E_\perp = F_{ab}\ell^ak^b \ \ \ , \ \ \ B_\perp = {}^*\!F_{ab}\ell^ak^b \ ,
\eea
where ${}^*\!F_{ab}$ is the dual of $F_{ab}$, namely 
${}^*\!F_{ab}=\frac{1}{2} \epsilon_{abcd}F^{cd}$ with 
$\epsilon_{abcd}$ the volume element of $g_{ab}$. 
Electric and magnetic charges can be expressed as (e.g. \cite{AshFaiKri00,Booth:2007wu})
\bea
\label{e:charges}
Q_{\mathrm{E}}= \frac{1}{4\pi}\int_{\cal S} E_\perp dS \ \ , \ \ 
Q_{\mathrm{M}}=\frac{1}{4\pi}\int_{\cal S} B_\perp dS \ .
\eea
When discussing the angular momentum in the presence of an electro-magnetic field, 
we add ${\cal L}_\eta A_a=0$ to the axisymmetry requirements (\ref{e:axisymmetry}).
Then, the following canonical notion of total angular momentum can be introduced 
on ${\cal S}$ \cite{Carte10,Simon:1984qb,Ashtekar:2001is, Dain:2011mv}
\bea
\label{e:angular_momentum}
J = J_{_{\mathrm{K}}} + J_{_{\mathrm{EM}}} =\frac{1}{8\pi}\int_{\cal S} \Omega_a^{(\ell)} \eta^a dS 
+ \frac{1}{4\pi}\int_S (A_a \eta^a) E_\perp dS \ .
\eea

\subsubsection{Yang-Mills fields}
Given a  Yang-Mills theory with Lie group $G$, the 
dynamical fields are given in terms of 
a 1-form $A_a$ evaluated on the Lie algebra ${\cal G}$ of $G$.
More properly,  $A_a$ is a connection on a principal $G-$bundle $P$ over the spacetime
$M$. Denoting the generators in ${\cal G}$ as $T_i$ and writing the Lie-algebra 
commutation relations as 
\begin{equation}
\label{e:commutation}
[T_i,T_j]=C^k_{ij}T_k \ ,
\end{equation}
the Cartan-Killing quadratic form on ${\cal G}$ is given by 
\begin{equation}
\label{e:CartanKilling}
{\rm k}_{ij}= C^k_{il}C^l_{jk} \ ,
\end{equation}
which is non-degenerate for semisimple Lie algebras.
For real compact Lie algebras, $\mathrm{k}_{ij}$ is 
non-degenerate and positive-definite 
(usually a basis $\{T_i\}$ of ${\cal G}$ such that
${\rm k}_{ij}=\delta_{ij}$ is employed). More generally,
the non-degenerate positive-definite character  of ${\rm k}_{ij}$
holds for Lie groups corresponding
to products of compact real Lie groups and  $U(1)$ factors.
Writing the Yang-Mills connection as ${\rm A}_a={A_a}^iT_i$, the Yang-Mills
tensor ${\rm F}_{ab}={F_{ab}}^iT_i$ is given by the curvature of ${\rm A}_a$, that
is ${\rm F}_{ab}=(d{\rm A})_{ab} + {\rm A}_a\wedge {\rm A}_b= 
\left(\nabla_a {A_b}^k - \nabla_b {A_a}^k + C^k_{ij} A^i_a A^j_b\right)T_k$. 
The Yang-Mills stress-energy tensor can be written
\bea
  \label{e:YM_Tab}
  T^{\mathrm{YM}}_{ab}= \frac{1}{4\pi}\mathrm{k}_{ij}\left({F_{ac}}^i
    {F_b{}^{c}}^j-\frac{1}{4}g_{ab} {F_{cd}}^i {F^{cd}}^j  \right).
\eea
We can define Yang-Mills electric and magnetic charges 
\cite{Chrusciel:1987jr,Sudarsky:1992ty,Ashtekar:2000hw} as
\bea
\label{e:YMcharges}
Q^{\mathrm{YM}}_{\mathrm{E}}=\frac{1}{4\pi}\int |{\rm E}^{\mathrm{YM}}_\perp|\;dS \ \ \ , \ \ \ 
Q^{\mathrm{YM}}_{\mathrm{M}} = \frac{1}{4\pi}\int |{\rm B}^{\mathrm{YM}}_\perp|\;dS \ ,
\eea
where
\bea
\label{e:YMelectricmagneticfields}
|{\rm E}^{\mathrm{YM}}_\perp|=\left[\left({F_{ab}}^ik^a\ell^b\right) 
\mathrm{k}_{ij} \left({F_{cd}}^jk^c\ell^d\right)\right]^{\frac{1}{2}} \ , \
|{\rm B}^{\mathrm{YM}}_\perp|=\left[\left({{}^*F_{ab}}^ik^a\ell^b\right) 
\mathrm{k}_{ij} \left({{}^*F_{cd}}^jk^c\ell^d\right)\right]^{\frac{1}{2}} \ .
\eea
Electromagnetic theory corresponds to the commutative
case $G=U(1)$. In particular, the  Yang-Mills principal 
fiber-bundle perspective sheds light on
the topological nature of the magnetic charge $Q_{\mathrm{M}}$, 
offering an understanding of magnetic monopoles as
associated with the non-triviality of the $U(1)$-bundle over $M$
(see e.g. \cite{WuYang75,Ryder:1980yb}), where $Q_{\mathrm{M}}$ is controlled
by the first Chern class of the $U(1)$-bundle.

\section{Stability of marginally outer trapped surfaces}
\label{s:stableMOTS}
The stability for marginally trapped surfaces is the crucial element
in the present discussion of the area inequalities.
This notion is extensively
reviewed in the contribution by M. Mars in this volume. We discuss the basic
elements here needed. 

\subsection{Basic definitions}

First, we choose conventionally $\ell^a$ as the {\em outgoing} 
null vector at ${\cal S}$, and refer to ${\cal S}$ as a 
{\em marginally outer trapped surface} (MOTS) if $\theta^{(\ell)}=0$.
Note that no condition is required on the ingoing expansion $\theta^{(k)}$.
The stability of MOTS is introduced in terms of the deformation operator $\delta_v$ 
on ${\cal S}$, that controls the infinitesimal variations of geometric objects 
defined on ${\cal S}$ under an infinitesimal deformation of the surface
 along a vector $v^a$ on ${\cal S}$ (here, $v^a$ will be always normal to ${\cal S}$). 
This operator $\delta_v$, discussed in detail in \cite{AndMarSim05,AndMarSim08} 
(see also M. Mars contribution and \cite{BooFai07,Cao:2010vj}),
is the analogue in the Lorentzian setting to the deformation operator employed 
in the discussion of minimal surfaces in Riemannian geometry.
We require ${\cal S}$ to be {\em stably outermost} in the 
sense introduced in \cite{AndMarSim05,AndMarSim08}
(see also  \cite{Hay94,Racz:2008tf}):
\begin{definition}
\label{d:stably_outermost}
{\em
Given a closed orientable marginally outer trapped surface ${\cal S}$
and a vector $v^a$ orthogonal to it, 
we will refer to ${\cal S}$ as {\em stably outermost with 
respect to the direction $v^a$} iff there exists a function
$\psi>0$ on ${\cal S}$ such that the variation
of $\theta^{(\ell)}$ with respect to $\psi v^a$
fulfills the condition $\delta_{\psi v} \theta^{(\ell)} \geq 0$.
}
\end{definition}
More specifically, we require ${\cal S}$ to be 
{\em spacetime stably outermost} 
\cite{Jaramillo:2011pg,Dain:2011kb}.
\begin{definition}
\label{d:spacetime_stably_outermost}
{\em 
A closed orientable marginally outer trapped surface ${\cal S}$ is referred
to as {\em spacetime stably outermost} if there exists 
an outgoing ($-k^a$-oriented) vector $X^a= \gamma \ell^a - 
\psi k^a$, with $\gamma\geq0$ and $\psi>0$, with respect to which 
${\cal S}$ is stably outermost: 
\bea
\label{e:stability_condition}
\delta_X \theta^{(\ell)} \geq 0.
\eea
If, in addition, $X^a$ (i.e. $\gamma$, $\psi$)
and $\Omega^{(\ell)}_a$ are axisymmetric,
we will refer to $\delta_X \theta^{(\ell)}\geq 0$ as an (axisymmetry-compatible) 
spacetime stably outermost condition.
}
\end{definition}
Alternatively, one could introduce the notion of stability for MOTS in terms of the non-negativity
of the principal eigenvalue $\lambda_v$ of the stability operator $L_v$ associated with
$\delta_v$, namely $L_v\psi= \delta_{\psi v}\theta^{(\ell)}$. Although  $L_v$ is not self-adjoint,
its principal eigenvalue (i.e. the eigenvalue with smallest real part) 
is indeed real. Then, the characterization in 
Definition \ref{d:stably_outermost} can be proved as a lemma~\cite{AndMarSim05,AndMarSim08}. 
This is the strategy followed in the contribution by M. Mars in this volume.

Finally, note that the characterization 
of MOTSs as spacetime stably outermost is independent of the choice
of future-oriented null normals $\ell^a$ and $k^a$. Indeed, 
given $f>0$, for $\ell'^a=f \ell^a$ and  $k'^a=f^{-1}k^a$ we can write
$X^a = \gamma \ell^a - \psi k^a= \gamma' \ell'^a - \psi' k'^a$
(with $\gamma'= f^{-1}\gamma\geq0$ and $\psi'= f \psi>0$), and it holds  
$\delta_X \theta^{(\ell')}=f \cdot \delta_X \theta^{(\ell)}>0$.

\subsection{Integral-inequality characterizations of MOTS stability}
\label{s:MOTSstability_characterization}
The first step in the proofs of area inequalities (\ref{e:inequality}) consists 
in casting condition (\ref{e:stability_condition}) as an integral
geometric inequality over ${\cal S}$.
Condition (\ref{e:stability_condition}) plays, for MOTS
in a Lorentzian context,  a role analogous to that of the stability condition 
for minimal surfaces in Riemannian geometry. 
In the Riemannian case this refers to the minimization of the
area of ${\cal S}$  with respect to arbitrary deformations along $\alpha s^a$,
where $s^a$ is the normal to ${\cal S}$ in a given 3-slice and $\alpha$ is an arbitrary
function on ${\cal S}$.
In contrast, the stability condition in Definition \ref{d:spacetime_stably_outermost} only
states the existence of a positive function $\psi$ (and, secondarily, of $\gamma\geq 0$).
The proof of area inequalities involving the angular momentum
requires writing (\ref{e:stability_condition}) as an integral inequality in terms
of arbitrary (axisymmetric) functions $\alpha$. 
The following lemma \cite{Jaramillo:2011pg} provides this\footnote{
Alternatively, one could start characterizing MOTS stability 
in terms of the principal eigenvalue $\lambda_X$. Then, the expression
of $\lambda_X$ in a Rayleigh-Ritz type characterization leads essentially to the
integral inequality. See M. Mars contribution, where the role of $\alpha$
is played by the function $u$.}

\vspace{0.1cm}
\noindent {\bf Lemma 1.~} 
{\em 
Given a closed orientable marginally outer trapped 
 surface ${\cal S}$ satisfying the 
spacetime stably outermost condition for an axisymmetric
$X^a$, then for all axisymmetric functions $\alpha$ on ${\cal S}$
\bea
\label{e:inequality_alpha}
&&\int_{\cal S} \left[D_a\alpha D^a\alpha + \frac{1}{2} \alpha^2 
\; {}^2\!R \right] dS \geq
 \\ 
&& \int_{\cal S} \left[ \alpha^2 \Omega^{(\eta)}_a  {\Omega^{(\eta)}}^a +
\alpha \beta \sigma^{(\ell)}_{ab} {\sigma^{(\ell)}}^{ab} 
+ G_{ab}\alpha\ell^a (\alpha k^b + \beta\ell^b) \right] dS \nn \ ,
\eea
where $\beta=\alpha\gamma/\psi$. If in addition we assume that the right hand
side in the inequality (\ref{e:inequality_alpha}) is not
identically zero, then ${\cal S}$ has a $S^2$ topology.
}

\vspace{0.1cm}
\begin{proof}
We basically follow the discussion in section 3.3. of 
\cite{Andersson:2010jv}
(cf. Th. 2.1 in \cite{Galloway:2005mf} for a similar reasoning,
essentially reducing a 
non time-symmetric case to a time-symmetric one). 
First, we evaluate $\delta_X \theta^{(\ell)}/\psi$  for
$X^a=\gamma \ell^a - \psi k^a$ in Definition 1,
with axisymmetric $\gamma$ and $\psi$.
For this we use (e.g. Eqs. (2.23) and (2.24) in \cite{BooFai07})
\bea
\delta_{\alpha \ell} \theta^{(\ell)} &=& \kappa^{(\alpha \ell)} \theta^{(\ell)}
- \alpha\left[\sigma^{(\ell)}_{ab} {\sigma^{(\ell)}}^{ab} + 
G_{ab} \ell^ak^b + \frac{1}{2}  \left(\theta^{(\ell)}\right)^2 \right] \nn \ ,
\\
\delta_{\beta k} \theta^{(\ell)} &=& \kappa^{(\beta k)} \theta^{(\ell)}
+ {}^2\!\Delta \beta - 2 \Omega^{(\ell)}_a  D^a\beta \nn \\
&&+\beta \left[ \Omega^{(\ell)}_a  {\Omega^{(\ell)}}^a 
- D^a  \Omega^{(\ell)}_a  -\frac{1}{2}{}^2\!R + G_{ab}k^a\ell^b 
-\theta^{(\ell)}\theta^{(k)} \right] \nn \ ,
\eea
where $\kappa^{(v)} = - v^a k^b \nabla_a \ell_b$.
Imposing $\theta^{(\ell)}=0$,  we can write for $X^a=\gamma \ell^a - \psi k^a$ 
\bea
\label{e:delta_X_theta}
\frac{1}{\psi}\delta_X\theta^{(\ell)}&=& 
- \frac{\gamma}{\psi} \left[\sigma^{(\ell)}_{ab} {\sigma^{(\ell)}}^{ab} 
+ G_{ab}\ell^a\ell^b \right] \nn \\
&&- {}^2\!\Delta \mathrm{ln}\psi -   
D_a\mathrm{ln}\psi D^a\mathrm{ln}\psi
+ 2 \Omega^{(\ell)}_a  D^a\mathrm{ln}\psi  \\
&&-\left[-D^a  \Omega^{(\ell)}_a 
+ \Omega^{(\ell)}_c  {\Omega^{(\ell)}}^c -\frac{1}{2}{}^2\!R + G_{ab}k^a\ell^b  \right] \nn \ .
\eea
We multiply  by $\alpha^2$ for arbitrary (axisymmetric) $\alpha$ 
and integrate on ${\cal S}$. Using 
$\int_{\cal S} \frac{\alpha^2}{\psi} \delta_X\theta^{(\ell)}dS\geq 0$ and
integrating by parts, we can write 
\bea
\label{e:int_delta_X_theta_1}
0\leq && \int_{\cal S}\alpha\beta \left[-\sigma^{(\ell)}_{ab} {\sigma^{(\ell)}}^{ab} 
- G_{ab}\ell^a\ell^b \right] dS\nn \\
&+&\int_{\cal S} \alpha^2 \left[-\Omega^{(\ell)}_a  {\Omega^{(\ell)}}^a 
+\frac{1}{2}{}^2\!R - G_{ab}k^a\ell^b \right] dS \nn \\ 
&+& \int_{\cal S}\left[2\alpha D_a\alpha D^a \mathrm{ln}\psi -   
\alpha^2 D_a\mathrm{ln}\psi D^a\mathrm{ln}\psi\right]
dS 
\nn \\
&+& \int_{\cal S}\left[2\alpha^2\Omega^{(\ell)}_a D^a\mathrm{ln}\psi 
- 2\alpha \Omega^{(\ell)}_a D^a\alpha\right]dS \ .
\eea
From the axisymmetry of $\alpha$ and $\psi$, ${\Omega^{(\eta)}}^aD_a\alpha=
{\Omega^{(\eta)}}^aD_a\psi=0$, and using (\ref{e:Omega_eta_xi})
\bea
\label{e:int_delta_X_theta_2}
0\leq && \int_{\cal S}\alpha\beta \left[-\sigma^{(\ell)}_{ab} {\sigma^{(\ell)}}^{ab} 
- G_{ab}\ell^a\ell^b \right] dS\nn \\
&+&\int_{\cal S} \alpha^2 \left[-\Omega^{(\eta)}_a  {\Omega^{(\eta)}}^a 
+\frac{1}{2}{}^2\!R - G_{ab}k^a\ell^b \right] dS \nn \\ 
&+& 
\int_{\cal S}\left[2 (D^a\alpha) (\alpha D_a \mathrm{ln}\psi -\alpha \Omega^{(\xi)}_a) 
\right. \\
&&\left.-(\alpha D_a\mathrm{ln}\psi -  \alpha \Omega^{(\xi)}_a)
(\alpha D^a\mathrm{ln}\psi -  \alpha {\Omega^{(\xi)}}^a)
\right]dS \ . \nn
\eea
Making use of the following Young's inequality in the last integral
\bea
\label{e:Young}
D^a\alpha D_a\alpha \geq 2 D^a\alpha (\alpha D_a \mathrm{ln}\psi -\alpha \Omega^{(\xi)}_a)
-|\alpha D\mathrm{ln}\psi -  \alpha \Omega^{(\xi)}|^2 
\eea
inequality (\ref{e:inequality_alpha}) follows for all axisymmetric $\alpha$.
Finally, if the right hand side  of (\ref{e:inequality_alpha}) does not vanish,
the sphericity of ${\cal S}$ follows by considering a constant 
$\alpha$ in (\ref{e:inequality_alpha}): it implies a positive value for the 
Euler characteristic of ${\cal S}$.
\end{proof}

The proof of the area-charge inequality, resulting from dropping
the angular momentum $J$ in (\ref{e:inequality}), requires neither a symmetry assumption
nor casting (\ref{e:stability_condition}) in terms of an arbitrary $\alpha$.
We use the following
lemma (slightly generalizing that in \cite{Dain:2011kb}).

\vspace{0.2cm}
\noindent {\bf Lemma 2.~} 
{\em 
  Given a closed orientable marginally outer trapped surface ${\cal S}$ satisfying the spacetime
  stably outermost condition, then the following inequality holds
\begin{equation}
\label{e:inequality_nonaxisym}
 \int_{\cal S} \left[ G_{ab}\ell^a   k^b 
+ N \left(\sigma^{(\ell)}_{ab} {\sigma^{(\ell)}}^{ab} + G_{ab}\ell^a \ell^b\right) \right]
 dS \leq  4\pi (1-g),  
\end{equation}
where $g$ is the genus of ${\cal S}$ and $N= \frac{\gamma}{\psi}\geq 0$. 
If in addition we assume that the left hand
side in the inequality (\ref{e:inequality_nonaxisym}) is non-negative and not
identically zero, then it follows that $g=0$ and hence ${\cal S}$ has the
$S^2$ topology.
}

\begin{proof}
The proof is slightly more simple than  that in Lemma 1.
We integrate directly expression (\ref{e:delta_X_theta}) over ${\cal S}$. 
On the  left hand side we use
the stability condition (\ref{e:stability_condition}). Divergence terms 
 in the right hand side integrate to zero and we rearrange terms as
\begin{equation}
  \label{eq:1}
-(D_a\mathrm{ln}\psi- \Omega^{(\ell)}_a )(D^a\mathrm{ln}\psi-
{\Omega^{(\ell)}}^a )=  
- D_a\mathrm{ln}\psi D^a\mathrm{ln}\psi
+ 2 \Omega^{(\ell)}_a  D^a\mathrm{ln}\psi - \Omega^{(\ell)}_c  {\Omega^{(\ell)}}^c,
\end{equation}
so that the integral is non-positive. From Gauss-Bonnet theorem, we write
\begin{equation}
  \label{eq:2}
  \int_S \frac{1}{2}{}^2\!R dS  =4\pi(1-g) \ \ .
\end{equation}
Collecting these observations, the inequality
(\ref{e:inequality_alpha}) follows. If the left hand side of the inequality
(\ref{e:inequality_alpha}) is non-negative it follows that $g$ can be $0$ or
$1$. If it is not identically zero then $g=0$ and hence ${\cal S}$ has the
$S^2$ topology.
\end{proof}

\subsection{Variants to the stably outermost condition}

\subsubsection{On an {\em averaged} outermost stably conditions for MOTS}
Inequalities (\ref{e:inequality_alpha}) and 
(\ref{e:inequality_nonaxisym}) do not require a {\em point-like} stability
condition. We could consider an (in principle weaker) 
{\em averaged} stability condition for MOTS.

\begin{definition}
\label{d:dipol-averaged}
{\em Given a closed orientable MOTS ${\cal S}$
we will refer to it as (dipole) {\em weight-averaged stably outermost} if
there exists an outgoing ($-k^a$-oriented) vector $x^a=\gamma \ell^a - k^a$, 
with $\gamma\geq0$ such that, for all functions $\alpha$ 
on ${\cal S}$, the variations
of $\theta^{(\ell)}$ with respect to $X^a = \alpha x^a$ fulfill the
integral condition 
\bea
\label{e:dipol_stability_condition}
\int_{\cal S} (X^a \ell_a) \delta_X \theta^{(\ell)} dS \geq 0 \ \ . 
\eea 
}
\end{definition}
Proofs of inequalities involving the angular momentum (cf. sections \ref{s:AJ} and \ref{s:AJQ})
could start from this averaged condition.
Note that $(X^a \ell_a)=\alpha=\mathrm{const}$ provides an
{\em averaged stably outermost} condition, the element needed in proving 
area-charge inequalities (cf. section \ref{s:AQ}).
Finally, a  ($2n$-moment) {\em weight-averaged stably outermost} condition could be introduced
as $\int_{\cal S} (X^a \ell_a)^n \;\delta_X \theta^{(\ell)} dS \geq 0$, for integers $n$.

\subsubsection{Towards axisymmetry relaxation}
\label{s:divergencefree}
Let $\eta^a$ be a divergence-free vector on ${\cal S}$, with
squared-norm $\eta=\eta^a\eta_a$ constant along itself, 
i.e.  $\eta^a D_a\eta=0$ (fulfilled, in particular, by Killing vectors).
As in \ref{s:axisymmetry}, we write
$q_{ab}=\frac{1}{\eta}\eta_a\eta_b + \xi_a\xi_b$,
with  $\xi^a\eta_a=\xi^a\ell_a=\xi^ak_a=0$, $\xi^a\xi_a=1$, and
$\Omega^{(\eta)}_a= \eta^b\Omega^{(\ell)}_b \eta_a/\eta$
and $\Omega^{(\xi)}_a= \xi^b\Omega^{(\ell)}_b \xi_a$, so relations (\ref{e:Omega_eta_xi})
hold. The geometric quantity 
\be
\label{e:qeta}
Q[\eta]=\frac{1}{4\pi}\int_{\cal S}  \frac{1}{\sqrt{\eta}}\Omega^{(\ell)}_a\eta^a dS \ \ ,
\ee
is well defined on ${\cal S}$ in the sense that: i)
it does not depend on the normalization
of the null normal $\ell^a$, and ii) there is no normalization ambiguity related 
to $\eta^a$. The first point follows 
from the transformation properties (\ref{e:null_transformations}) of $\Omega^{(\ell)}_a$ 
under $\ell'^a =f \ell^a$,
together with the divergence-free character
of $\hat{\eta}^a=\eta^a/\sqrt{\eta}$, i.e.
$D_a\left(\hat{\eta}^a\right)
= \frac{1}{\sqrt{\eta}}D_a\eta^a - \frac{1}{2\eta\sqrt{\eta}}\eta^a D_a\eta =0$. 
Regarding the second point, $Q[\eta]$ is defined in terms 
of $\hat{\eta}^a$, with $\hat{\eta}^a\hat{\eta}_a=1$.
This is the analogue of the $2\pi$-orbit normalization for
axial Killing vectors in expressions (\ref{e:Komar_angular_momentum}) and
(\ref{e:angular_momentum}) for 
the angular momentum (note that $\eta^a$ needs not to be axial).
We can then adapt the MOTS stability condition: 

\begin{definition}
{\em 
Given a closed orientable marginally outer trapped surface ${\cal S}$ and 
a divergence free vector $\eta^a$ on it,
${\cal S}$ is said to be $\eta^a$-compatible spacetime stably outermost
if there exists an outgoing ($-k^a$-oriented) vector $X^a= -\psi k^a$, with 
$\psi>0$ and $\eta^aD_a\psi=0$, such that the variation
of $\theta^{(\ell)}$ with respect to $X^a$ fulfills the condition
$\delta_X \theta^{(\ell)} \geq 0$.}
\end{definition}

The following lemma holds.
\vskip 0.2cm
{\em 
\noindent {\bf Lemma 3.~} 
\label{l:eta_stable}
Given a closed orientable MOTS ${\cal S}$ satisfying the $\eta^a$-compatible 
spacetime stably outermost condition for 
$X^a$, then for all $\alpha$ such that $\eta^aD_a\alpha=0$, it holds
\bea
\label{e:inequality_eta_alpha}
\int_{\cal S} \left[D_a\alpha D^a\alpha + \frac{1}{2} \alpha^2 
\; {}^2\!R \right] dS \geq
\int_{\cal S} \left[ \alpha^2 \left(\Omega^{(\eta)}_a  {\Omega^{(\eta)}}^a +
G_{ab}\ell^a k^b\right) \right] dS  \ ,
\eea
}
\begin{proof}
The proof proceeds exactly as in Lemma 1.
It is straightforward to generalize it for $X^a= \gamma \ell^a - 
\psi k^a$, $\gamma \geq 0$, so that the shear and the $G_{ab}\ell^a \ell^b$ terms
are incorporated. 
\end{proof}

\section{The area-angular momentum inequality}
\label{s:AJ}
We first state the main result in this section (see \cite{Jaramillo:2011pg} for 
further details):

\vskip 0.2cm
\noindent {\bf Theorem 1~} (cf. Ref. \cite{Jaramillo:2011pg}).
{\em Given an axisymmetric closed orientable marginally outer trapped
surface ${\cal S}$ satisfying the (axisymmetry-compatible)
spacetime stably outermost condition, in a spacetime  with non-negative cosmological constant
and fulfilling the dominant energy condition, 
it holds the inequality
\bea
\label{e:inequality_AJ}
A \geq 8\pi |J| \ \ ,
\eea
where $A$ and $J$ are the area
and gravitational 
(Komar) angular momentum of ${\cal S}$. If equality holds, then 
${\cal S}$ has the geometry of an extreme Kerr throat sphere and, in 
addition, if the vector $X^a$
in the stability condition can be found to be spacelike then
${\cal S}$  is a section of a non-expanding horizon.
}

\vskip 0.2cm
The proof of the area-angular momentum inequality (\ref{e:inequality_AJ})
has two parts.
The first one is purely geometric and provides the lower bound on the area $A$
\bea
\label{e:area_bound}
A\geq 4\pi e^{\frac{{\cal M} -8}{8}} \ \ ,
\eea
where ${\cal M}$ is a functional on the sphere geometry.
The second part solves a variational problem, subject to the
constraint of keeping constant the a priori given angular momentum $J$. 
In particular, it is shown \cite{Acena:2010ws,GabachClement:2011kz,Dain:2011pi} 
the existence of a minimum 
\bea
\label{e:Mminimum}
{\cal M}\geq {\cal M}_0 \  \ , \ \ \ \ \hbox{ under the constraint }   J  \hbox{ fixed} 
\eea
such that the evaluation of (\ref{e:area_bound}) on ${\cal M}_0$ leads to 
inequality (\ref{e:inequality_AJ}). Moreover the minimizer is unique, this
leading to a rigidity result.
We focus here on the first geometric part and refer the reader 
to the proper references on the variational part \cite{Acena:2010ws,GabachClement:2011kz,Dain:2011pi}.

\begin{proof}
First, we consider an axisymmetric stably outermost MOTS and apply the result in Lemma 1,
namely, we consider inequality  (\ref{e:inequality_alpha}) where 
we disregard the positive-definite gravitational radiation 
shear squared term. Imposing Einstein equation, we also disregard 
the cosmological constant and matter terms under the assumption of non-negative
 cosmological constant $\Lambda\geq 0$
and the dominant energy condition
(note that $\alpha k^b + \beta\ell^b$ is a non-spacelike vector).
Therefore
\be
\label{e:geom_inequality_eta}
\int_{\cal S} \left[D_a\alpha D^a\alpha + \frac{1}{2} \alpha^2 
\; {}^2\!R \right] dS\geq \int_{\cal S} \alpha^2 \Omega^{(\eta)}_a  {\Omega^{(\eta)}}^a dS.
\ee
Second,  we express this inequality in terms of certain  
potentials for the geometry of ${\cal S}$.
Assuming a non-vanishing right hand side in (\ref{e:geom_inequality_eta})
(otherwise (\ref{e:area_bound}) is trivial),  ${\cal S}$ has a spherical 
topology. On an axisymmetric sphere  we can always write \cite{AshEngPaw04}
\bea
\label{e:q_ab}
ds^2=q_{ab}dx^a dx^b = e^\sigma \left(e^{2q} d\theta^2 + 
\mathrm{sin}^2\theta d\varphi^2 \right) \ ,
\eea
with $\sigma$ and $q$  functions on $\theta$ satisfying $\sigma+q=c$, where $c$ is a constant. 
Then $dS=e^c dS_0$, with 
$dS_0= \mathrm{sin}\theta d\theta d\varphi$. In addition,
the squared norm $\eta$ of the axial Killing vector $\eta^a=(\partial_\varphi)^a$ is given by 
\bea
\label{e:eta}
\eta = e^\sigma \mathrm{sin}^2\theta \ .
\eea
Choosing $\alpha=e^{c-\sigma/2}$ \cite{Dain:2011pi}, the evaluation of
the left hand side in (\ref{e:geom_inequality_eta})
results in 
\bea
\label{e:analytic_lhs}
&&\int_{\cal S} \left[D_a\alpha D^a\alpha + \frac{1}{2} \alpha^2 
\; {}^2\!R \right] dS
= e^c\left[4\pi(c+1)-\int_{\cal S}\left(\sigma+\frac{1}{4}
\left(\frac{d\sigma}{d\theta}\right)^2\right)
dS_0\right]. 
\eea
To evaluate the right hand side in (\ref{e:geom_inequality_eta})
we note that, due to the $S^2$ topology of ${\cal S}$, we can always express
$\Omega^{(\ell)}_a$ in terms of a divergence-free and an exact
form
\bea
\label{e:Omega_ell}
\Omega^{(\ell)}_a = \epsilon_{ab}D^b \tilde{\omega}
+D_a\lambda \ ,
\eea
with $\tilde{\omega}$ and $\lambda$ fixed up to a constant.
From the axisymmetry of $q_{ab}$ and $\Omega^{(\ell)}_a$ (functions
$\tilde{\omega}$ and $\lambda$ are then axially symmetric) it follows
$\Omega^{(\eta)}_a= \epsilon_{ab}D^b \tilde{\omega}$
and $\Omega^{(\xi)}_a =D_a\lambda$. Before proceeding further, 
we evaluate the angular momentum $J$. Writing
$\eta^a \Omega^{(\ell)}_a = \eta^a \Omega^{(\eta)}_a = \epsilon_{ab}\eta^a 
D^b\tilde{\omega}$ and expressing $\xi^a$
as $\xi_b=\eta^{-1/2} \epsilon_{ab}\eta^a$, we have
\be
\label{e:eta_Omega}
   \Omega^{(\ell)}_a  \eta^a=  \eta^{1/2} \xi^a D_a\tilde \omega \ .
\ee
Plugging this into Eq. (\ref{e:Komar_angular_momentum})
(or (\ref{e:angular_momentum}), since $J_{\mathrm {EM}} =0$) and using 
(\ref{e:q_ab}) we find
\be
\label{eq:J_omega}
  J=\frac{1}{8}\int_{0}^\pi  2\eta\frac{d\tilde\omega}{d\theta}
   =\frac{1}{8}\int_{0}^\pi  \frac{d\bar \omega}{d\theta}
   =\frac{1}{8} \left[\bar \omega(\pi)-\bar \omega(0) \right] \ ,
\ee
where we have introduced the potential $\bar\omega$ as
$d\bar\omega/d\theta \equiv (2\eta) d\tilde\omega/d\theta$.
The use of $\bar\omega$, rather than $\tilde\omega$,
permits to control directly the angular momentum in terms
of the values of  $\bar\omega$ at the axis.
This is crucial to implement the constraint $J=\mathrm{const}$
in the variational problem. 
We use $\bar{\omega}$  in the following, rather than $\tilde\omega$.
Further geometric intuition is gained
by noting that, if the axial vector $\eta^a$ on ${\cal S}$ extends 
to a spacetime neighborhood of ${\cal S}$ 
(something not needed in the present discussion),
we can define the {\em twist} vector of $\eta^a$ as
$\omega_a= \epsilon_{abcd}\eta^b \nabla^c\eta^d$ and the relation
$\xi^a \omega_a=\xi^aD_a\bar{\omega}$ holds. In the vacuum case,
a spacetime twist potential $\omega$ satisfying $\omega_a = \nabla_a\omega$
can be defined, so that $\bar{\omega}$ and $\omega$ coincide on
${\cal S}$ up to a
constant. 
Note however that $\bar{\omega}$ on ${\cal S}$ can be defined always, even in the
presence of matter.

From Eqs. (\ref{e:Omega_ell}) and (\ref{e:q_ab}) and the adopted choice for 
$\alpha$, we have
\be
\label{e:rhs_ineq}
\alpha^2 \Omega^{(\eta)}_a  {\Omega^{(\eta)}}^a 
=\frac{\alpha^2}{4\eta^2} D_a\bar{\omega} D^a\bar{\omega} = 
\frac{1}{4\eta^{2}}\left(\frac{d\bar{\omega}}{d\theta}\right)^{2} \ .
\ee
Plugging this into  (\ref{e:geom_inequality_eta}) and using (\ref{e:analytic_lhs})
we get
\bea
\label{e:ineqM}
8(c+1)\geq {\cal M}[\sigma, \bar{\omega}] \ ,
\eea
with
\bea
\label{e:functional}
{\cal M}[\sigma, \bar{\omega}]= \frac{1}{2\pi} \int_{\cal S}
\left[\left(\frac{d\sigma}{d\theta}\right)^2 + 
4\sigma
+ \frac{1}{\eta^2} \left(\frac{d\bar{\omega}}{d\theta}\right)^2
\right]dS_0 \ .
\eea
Using these expressions and $A=4\pi e^c$ leads to inequality (\ref{e:area_bound}).
This completes the first stage in the proof. In a second stage, by
solving the variational problem defined by ${\cal M}[\sigma, \bar{\omega}]$
with $J$ constant as a constraint, one can prove \cite{Acena:2010ws,GabachClement:2011kz}
\bea
\label{e:MgeqM0_v1}
{\cal M}\geq {\cal M}_0 = 8\ln(2|J|) + 8 \ .
\eea
 This, namely $ e^{({\cal M} -8)/8}\geq 2|J|$, together
with (\ref{e:area_bound}) leads to area-angular momentum inequality
(\ref{e:inequality_AJ}). Actually, the only minimizer for 
${\cal M}_0$ in (\ref{e:MgeqM0_v1}) is  extremal Kerr, this leading to a 
rigidity result \cite{Dain:2011pi,Jaramillo:2011pg}: if equality in 
(\ref{e:inequality_AJ}) holds, first, the intrinsic geometry of ${\cal S}$
is that of an extreme Kerr throat sphere \cite{Dain:2010qr}
and, second, the vanishing of the positive-definite terms 
in (\ref{e:inequality_alpha}) implies,  
for spacelike $X^a$ in (\ref{e:stability_condition}),
the vanishing of the shear $\sigma^{(\ell)}_{ab}$
so that ${\cal S}$ is an {\em instantaneous }
(non-expanding) isolated horizon \cite{AshKri04}.

\end{proof}

\section{The area-charge inequality. Generalizations}
\label{s:AQ}
Remarkably, if we drop the angular momentum from the
inequality (\ref{e:inequality}) , the resulting area-charge inequality requires neither
the use of a variational principle nor the assumption of any symmetry.

\vskip 0.2cm
\noindent {\bf Theorem 2~} (cf. Ref. \cite{Dain:2011kb}).
{\em
Given an orientable closed orientable marginally outer trapped surface ${\cal S}$ satisfying
the spacetime stably outermost condition, in a spacetime which satisfies 
Einstein
equations, with non-negative cosmological constant $\Lambda$ and
such that the non-electromagnetic matter fields $T_{ab}$ satisfy the dominant
energy condition, then it holds 
\begin{equation}
\label{e:inequality_AQ}
A \geq 4\pi \left(Q_{\mathrm{E}}^2+Q_{\mathrm{M}}^2\right) ,
\end{equation}
where $A$,  $Q_{\mathrm{E}}$ and $Q_{\mathrm{M}}$ are the area, electric and magnetic charges of
${\cal S}$.
}
\vskip 0.2cm

\begin{proof}
We start from Lemma 2 and use inequality (\ref{e:inequality_nonaxisym}) and Einstein equations
(\ref{e:Einstein_eq}). Since the vector $k^a+\gamma/\psi \ell^a$ is timelike or null,
  using that the tensor $T_{ab}$ satisfies the dominant energy condition
(and in particular the null energy condition),
  that $\Lambda$ is non-negative and the term proportional to $N$ is definite-positive,
we get from (\ref{e:inequality_nonaxisym})   
\begin{equation}
  \label{eq:10}
  8\pi  \int_{\cal S}  T^{\mathrm{EM}}_{ab}\ell^a k^b   dS  \leq 4\pi(1-g). 
\end{equation}
The term $T^{\mathrm{EM}}_{ab}\ell^a k^b$ can be written as 
\bea
\label{e:T_EM_lk}
    T^{\mathrm{EM}}_{ab}\ell^ak^b = \frac{1}{8\pi} 
    \left[\left(\ell^a k^b F_{ab}\right)^2 
    +\left(\ell^a k^b {}^*\!F_{ab}\right)^2\right].
\eea
This result is purely algebraic, something crucial for the later generalization
to Yang-Mills fields. 
To derive (\ref{e:T_EM_lk}) we use the decomposition (\ref{e:metric_q}) for $g_{ab}$ 
and calculate
\bea
    \label{e:FF1}
    F_{ab}F^{ab}=-2\left( \ell^a k^b F_{ab} \right)^2-4q^{ab} k^c F_{ac} \ell^d
    F_{bd}+ F_{ab} F_{cd}q^{ac}q^{bd},
\eea
and
\bea
  \label{e:FF2}
  \ell^a  k^c F_{ab} F_{c}{}^b= \left( \ell^a k^b F_{ab} \right)^2 + q^{ab} k^c
  F_{ac} \ell^d F_{bd}.
\eea
Noting that the pull-back of $F_{ab}$ on the surface ${\cal S}$ is proportional to
the volume element $\epsilon_{ab}$ of the surface ${\cal S}$, we can evaluate
$F_{ab} F_{cd}q^{ac}q^{bd}$ and $\left(\epsilon^{ab}F_{ab}\right)^2$ to obtain
\bea
  \label{e:FF3}
   F_{ab} F_{cd}q^{ac}q^{bd} = \frac{1}{2} \left(\epsilon^{ab}F_{ab}\right)^2
   = 2 \left({}^*\!F_{ab}\ell^ak^b\right)^2 \ ,
\eea
where the identity ${}^*\!F_{ab}\ell^ak^b = \frac{1}{2} F_{ab}\epsilon^{ab}$
follows from the relation $\epsilon_{ab}= \epsilon_{abcd}\ell^c k^d$.  
Inserting these expressions into Eq. (\ref{e:EMstressenergy})
 we obtain (\ref{e:T_EM_lk}). Then, using relation (\ref{e:T_EM_lk}) into inequality (\ref{eq:10})
we get
\bea
  \label{e:FFg}
\int_{\cal S}   \left[\left(\ell^a k^b F_{ab}\right)^2 
    +\left(\ell^a k^b {}^*\!F_{ab}\right)^2\right]dS \leq 4\pi(1-g).
\eea
If the left hand side is identically zero then
the charges are zero and the inequality (\ref{e:inequality_AQ}) is trivial. 
We assume that it is not zero at some point and hence we have $g=0$.
To bound the left hand side of inequality (\ref{e:FFg}) we use H\"older
inequality on ${\cal S}$. For
integrable functions $f$ and $h$, H\"older inequality is given by
\begin{equation}
  \label{eq:12}
  \int_{{\cal S}}  fh dS\leq  \left(\int_{\cal S}  f^2 dS\right)^{1/2}
  \left(\int_{{\cal S}}  h^2 dS\right)^{1/2}.
\end{equation}
If we take $h=1$, then we obtain
\begin{equation}
  \label{eq:13}
  \int_{{\cal S}}  f dS\leq  \left(\int_{{\cal S}}  f^2 dS\right)^{1/2}  A^{1/2}.
\end{equation}
where $A$ is the area of ${\cal S}$. 
Using this inequality in (\ref{e:FFg}) we finally obtain
\begin{equation}
  \label{eq:14}
  A^{-1}\left[ \left (\int_{\cal S} \ell^a k^b F_{ab} dS \right)^2
      + \left (\int_{\cal S} \ell^a k^b {}^*\!F_{ab} dS \right)^2 \right]
\leq 4\pi. 
\end{equation}
Finally, we use the expression of the charges (\ref{e:charges}) to express the
 left-hand-side of (\ref{eq:14})  in terms of $Q_{\mathrm{E}}$ and  $Q_{\mathrm{M}}$.
Hence the inequality (\ref{e:inequality_AQ}) follows. 

\end{proof}

\subsection{Yang-Mills charges}
\label{s:AQYM}
The derivation of the area-charge inequality (\ref{e:inequality_AQ})
does not involve Maxwell equations, 
only the algebraic form of the electromagnetic stress-energy tensor 
(\ref{e:EMstressenergy}) is used. Given the similar structure of
the Yang-Mills stress-energy tensor (\ref{e:YM_Tab}), the result 
generalizes to include Yang-Mills charges (\ref{e:YMcharges}), for compact Lie groups.

\vskip 0.2cm
\noindent {\bf Corollary 1.~} 
{\em 
Under the conditions of Theorem 2,
for a Yang-Mills theory with compact simple Lie group 
$G$ (more generally with $G$ given by a product of 
compact simple Lie groups and $U(1)$ factors)
it holds the inequality
\be
\label{e:YM_ineq}
A\geq 4\pi\left[\left(Q^{\mathrm{YM}}_{\mathrm{E}}\right)^2 + \left(Q^{\mathrm{YM}}_{\mathrm{M}}\right)^2 \right] \ .
\ee
}
\begin{proof}
Proceeding exactly as in Theorem 2 and writing
\bea
    \label{e:T_F_Ftilde}
    T^{\mathrm{YM}}_{ab}\ell^ak^b = \frac{1}{8\pi}\mathrm{k}_{ij} 
    \left[\left(\ell^a k^b {F_{ab}}^i\right)\left(\ell^c k^d {F_{cd}}^j\right) 
    +\left(\ell^a k^b {{}^*\!F_{ab}}^i\right)
     \left(\ell^c k^c {{}^*\!F_{cd}}^j\right)\right].
\eea
we derive the analogue of inequality (\ref{e:FFg})
\bea
\label{e:YM_4piineq}
4\pi\geq \int_{\cal S} 
\left[\left(\ell^a k^b {F_{ab}}^i\right)\mathrm{k}_{ij}
\left(\ell^c k^d {F_{cd}}^j\right) 
+\left(\ell^a k^b {{}^*\!F_{ab}}^i\right)\mathrm{k}_{ij} 
     \left(\ell^c k^c {{}^*\!F_{cd}}^j\right)\right]dS \ .
\eea
In this case, we can write the form (\ref{eq:13}) of H\"older inequality 
as
\begin{equation}
  \label{eq:Holder_YM}
  \int_{\cal S} \mathrm{k}_{ij}V^iV^j \geq \frac{1}{A}
  \left(\int_{\cal S}  \left(\mathrm{k}_{ij}V^iV^j\right)^{\frac{1}{2}} dS 
\right)^2 \ ,
\end{equation}
for compact Lie algebras, for which $\mathrm{k}_{ij}$ in (\ref{e:CartanKilling}) 
is definite-positive
[just take $f^2=\mathrm{k}_{ij}V^iV^j\geq0$ in (\ref{eq:13})].
Using inequality (\ref{eq:Holder_YM}) in (\ref{e:YM_4piineq}) leads to 
inequality (\ref{e:YM_ineq}).

\end{proof}

\subsection{Further generalizations}
The area-charge inequality can be extended to incorporate the
quantity $Q[\eta]$ in (\ref{e:qeta}).

\vskip 0.2cm
\noindent {\bf Corollary 2.~} 
{\em
Under the conditions of Lemma 3,
the following inequality holds
\bea
\label{e:AQ_eta}
A\geq 4\pi Q[\eta]^2 \ .
\eea
}
\begin{proof}
Starting from inequality (\ref{e:inequality_eta_alpha}),
choose $\alpha=1$ and drop the electromagnetic or Yang-Mills components 
(it is straightforward to include them). Then we can write
\bea
4\pi\geq \int_{\cal S} \Omega^{(\eta)}_a  {\Omega^{(\eta)}}^a dS
= \int_{\cal S}\frac{1}{\eta} \left(\Omega^{(\ell)}_a \eta^a\right)^2dS=
\int_{\cal S}  \left(\frac{1}{\sqrt{\eta}}\Omega^{(\ell)}_a \eta^a\right)^2dS \ .
\eea
Using again inequality (\ref{eq:13}), now with 
$f=\frac{1}{\sqrt{\eta}}\Omega^{(\ell)}_a \eta^a$, we obtain
\bea
 \int_{\cal S}\left(\frac{1}{\sqrt{\eta}}\Omega^{(\ell)}_a \eta^a\right)^2dS\geq
\frac{1}{A}\left(\int_{\cal S}\frac{1}{\sqrt{\eta}}
\Omega^{(\ell)}_a \eta^a dS \right)^2 \ ,
\eea
from which inequality (\ref{e:AQ_eta}) follows when using expression 
(\ref{e:qeta}) for $Q[\eta]$.
\end{proof}
Two remarks are in order.
First, inequality (\ref{e:AQ_eta}) does not reduce to
the area-angular momentum inequality (\ref{e:inequality_AJ}), even if $\eta^a$ is an axial Killing
vector. Even in this case, the quantity $Q[\eta]$ is not an angular momentum
due to the $1/\sqrt{\eta}$ factor (this is easily seen on dimensional grounds).
However, whenever existing,  $Q[\eta]$ is a geometric quantity providing a
non-trivial lower bound for the area. 
Second, the area-charge geometric inequalities (\ref{e:inequality_AQ}),  (\ref{e:YM_ineq}) 
and (\ref{e:AQ_eta}) can be collected in the more general form
\bea
\label{e:general}
A\geq 4\pi \left[Q_{\mathrm{E}}^2 + Q_{\mathrm{M}}^2 + 
\left(Q^{\mathrm{YM}}_{\mathrm{E}}\right)^2 + \left(Q^{\mathrm{YM}}_{\mathrm{M}}\right)^2 + Q[\eta]^2
\right] \ ,
\eea
assuming that the individual terms make sense.

\subsubsection{The Cosmological constant and stability operator eigenvalue}
\label{s:AQlambda}
The area-charge inequality has been extended in Ref. \cite{Simon:2011zf}
to include the cosmological constant $\Lambda$ and the principal eigenvalue
$\lambda$ of the stability operator associated with 
the deformation operator $\delta$ ($\lambda$ is a real number \cite{AndMarSim05,AndMarSim08}). 
The inequality reads
\bea
\label{e:A-Q-Lambda}
\Lambda^* A^2-4\pi(1-g) A + (4\pi)^2\sum_i Q_i^2 \leq 0 \ ,
\eea 
where  $\Lambda^*\equiv \Lambda + \lambda$  and $Q_i$ correspond to $Q_{\mathrm{E}}$, $Q_{\mathrm{M}}$, $Q^{\mathrm{YM}}_{\mathrm{E}}$, $Q^{\mathrm{YM}}_{\mathrm{M}}$ and $Q[\eta]$.
The previous inequality (\ref{e:general}) follows from the 
stability condition $\Lambda^*>0$ and $g=0$.
We highlight the remarkable fact that the cosmological constant and the 
principal eigenvalue enter formally in exactly the same manner. This suggests
the possibility of linking global and quasi-local notions of stability in 
black hole spacetime geometries.

\subsubsection{Energy flux terms}
\label{s:AQflux}
From a physical perspective, it is suggestive to rewrite the previous inequality 
(\ref{e:A-Q-Lambda}) without dropping neither the matter terms nor 
the piece proportional to $N$ in
(\ref{e:inequality_nonaxisym}). Following \cite{AshKri02,AshKri03} we define
${\cal F}_{\mathrm{grav}}\equiv \frac{1}{16\pi}\int_{\cal S} N \sigma^{(\ell)}_{ab} {\sigma^{(\ell)}}^{ab} dS$
as the instantaneous flux of (transverse \cite{Hay04,Hayward04}) gravitational radiation 
measured by an (Eulerian) observer associated 
with a foliation with lapse function $N$. Expressing the flux of matter energy
as ${\cal F}_{\mathrm{matter}}\equiv \int_{\cal S} T^{\mathrm{M}}_{ab} \ell^a t^b dS$ (with $t^a=k^b + N \ell^b$ 
a timelike vector) and the electromagnetic Poynting flux as 
${\cal F}_{\mathrm EM} = \int_{\cal S} N T^{\mathrm{EM}}_{ab} \ell^a \ell^b dS$, we write (with $g=0$)
\bea
\label{e:inequality_flux}
\frac{1}{2} \geq \frac{\Lambda^*}{2} \left(\frac{A}{4\pi}\right) + 
\frac{1}{2}\left(\frac{4\pi}{A}\right)\sum_i Q_i^2 
+ {\cal F}_{\mathrm{EM}} + {\cal F}_{\mathrm{matter}} + 2 {\cal F}_{\mathrm{grav}}
\eea 
This emphasizes the role of integral inequalities (\ref{e:inequality_alpha}), 
(\ref{e:inequality_nonaxisym})
and (\ref{e:inequality_eta_alpha}) in Lemmas 1, 2 and 3
as {\em energy flux} inequalities.
In particular, flux inequality  (\ref{e:inequality_flux}) indicates that the instantaneous 
flux of energy into a stable black hole horizon is bounded from above 
so that it cannot be arbitrarily large.

\section{The area-angular momentum-charge inequality}
\label{s:AJQ}
After discussing the area-angular momentum and area-charge inequalities,
we address now the inequality incorporating all relevant quantities in
Einstein-Maxwell theory.

\vspace{0.2cm}
\noindent {\bf Theorem 3~} (cf. Refs. \cite{Clement:2011tq,GabJarRei11}). 
{\em Given an axisymmetric closed orientable marginally outer trapped
surface ${\cal S}$ satisfying the (axisymmetry-compatible)
spacetime stably outermost condition, in a spacetime 
with non-negative cosmological constant and matter
content fulfilling the dominant energy condition, 
it holds the inequality
\bea
\label{e:inequality_AJQ}
\left(A/(4\pi)\right)^2\geq (2 J)^2 + (Q_{\mathrm{E}}^2 + Q_{\mathrm{M}}^2)^2
\eea
where $A$ is the area of ${\cal S}$ and $J$, $Q_{\mathrm{E}}$ and $Q_{\mathrm{M}}$ are, respectively,
the total (gravitational and electromagnetic) angular momentum, the electric
and the magnetic charges associated with ${\cal S}$.  If equality holds, then 
${\cal S}$ has the geometry of an extreme Kerr-Newman throat sphere and, in 
addition, if vector $X^a$
in the stability condition can be found to be spacelike then
${\cal S}$  is a section of a non-expanding horizon.
}

\begin{proof}
The proof \cite{Clement:2011tq,GabJarRei11} 
follows the steps in Theorem 1, namely with a first stage in which
a lower bound (\ref{e:area_bound}) on the area is derived, followed by the
resolution of a variational
problem under the constraints of keeping  $J$, $Q_{\mathrm{E}}$ and $Q_{\mathrm{M}}$ fixed.

First, starting from inequality (\ref{e:inequality_alpha}), proceeding then as in 
the derivation of (\ref{e:geom_inequality_eta}) and using relations 
(\ref{e:T_EM_lk}) and (\ref{e:E_B_normal}),  we obtain
\bea
\label{e:geom_inequality_alpha_EB}
\int_{\cal S} \left[|D \alpha|^2 + \frac{1}{2} \alpha^2 
\; {}^2\!R \right] dS 
\geq \int_{\cal S} \alpha^2 \left[|\Omega^{(\eta)}|^2 
+ (E_\perp^2 + B_\perp^2)\right]dS \ . 
\eea
From this expression, contact can be made \cite{GabachClement:2011kz} 
with the proof in \cite{Hennig:2008zy}
to establish inequality (\ref{e:inequality_AJQ}) for vanishing  $Q_{\mathrm{M}}$. 
Here, we rather follow \cite{Clement:2011tq,GabJarRei11} the strategy in section \ref{s:AJ}.
In order to identify the relevant action functional ${\cal M}$ for the variational problem, 
in particular its dependence on appropriate potentials permitting to control 
the constraints on $J$, $Q_{\mathrm{E}}$ and $Q_{\mathrm{M}}$, 
we adopt again a coordinate system (\ref{e:q_ab}) on the axisymmetric sphere and use 
the decomposition (\ref{e:Omega_ell}) introducing the potential $\tilde{\omega}$.
From expressions (\ref{e:charges}) for $Q_{\mathrm{E}}$ and $Q_{\mathrm{M}}$ and 
(\ref{e:angular_momentum})
for $J$, we write (see details in \cite{GabJarRei11})
\bea
\label{e:Q_EQ_MJ}
Q_{\mathrm{E}}&=& \frac{1}{2} \int_0^\pi E_\perp e^c \sin\theta d\theta=
\frac{1}{2}[\psi(\pi)- \psi(0)] \nn \\
Q_{\mathrm{M}}&=& \frac{1}{4\pi} \int_0^\pi \frac{dA_\varphi}{d\theta}  d\theta =
\frac{1}{2}[\chi(\pi)- \chi(0)] \nn \\
J&=&\frac{1}{8} \int_0^\pi \left(2\eta\frac{d\tilde\omega}{d\theta}
+2\chi\frac{d\psi}{d\theta}-2\psi\frac{d\chi}{d\theta}\right) = 
\frac{1}{8}[\omega(\pi)- \omega(0)] \ ,
\eea
where we have introduced the new potentials $\omega$, $\chi$ and $\psi$ on ${\cal S}$
\bea
\label{e:omega_chi_psi}
\frac{d\psi}{d\theta}&=&E_\perp e^c\sin\theta \ \ \ \ , \ \ \ \ \chi=A_\varphi \ , \nn \\
\frac{d\omega}{d\theta}&=& 2\eta\frac{d\tilde\omega}{d\theta}
+2\chi\frac{d\psi}{d\theta}-2\psi\frac{d\chi}{d\theta} =
\frac{d\bar\omega}{d\theta}
+2\chi\frac{d\psi}{d\theta}-2\psi\frac{d\chi}{d\theta} \ .
\eea
Therefore fixing $\omega$, $\chi$ and $\psi$ on
the axis does control the values of $Q_{\mathrm{E}}$ and $Q_{\mathrm{M}}$ and $J$ in the
variational problem. Using these potentials in (\ref{e:geom_inequality_alpha_EB}), 
with $\alpha=e^{c-\sigma/2}$, we get
\bea
\label{e:ineqM2}
8(c+1)\geq {\cal M}[\sigma, \omega, E_\perp, A_\varphi] \ ,
\eea
where
\bea
\label{e:functional_v2}
&&\!\!\! {\cal M}[\sigma, \omega, \psi, \chi]=
\frac{1}{2\pi}\int_{\cal S}\left[ 4\sigma+|D\sigma|^2 \right. \\
&&\left. + \frac{|D\omega-2\chi D\psi+2\psi D\chi|^2}{\eta^2}
+\frac{4}{\eta}(|D\psi|^2+|D\chi|^2) \right]dS_0 \nn \ ,
\eea
from which an inequality (\ref{e:area_bound}) is recovered by using $A=4\pi e^c= 4\pi e^{\sigma(0)}$.
The proof of the area-charge-angular momentum inequality (\ref{e:inequality_AJQ}) is completed by showing that 
\bea
\label{e:Mminimum}
{\cal M}\geq {\cal M}_0  = 8\ln\sqrt{(2J)^2 + (Q_{\mathrm{E}}^2+ Q_{\mathrm{M}}^2)^2} + 8 \ , 
\eea
under the constraint of keeping $J$, $Q_{\mathrm{E}}$ and $Q_{\mathrm{M}}$ fixed.  
Here ${\cal M}_0$ corresponds to the evaluation of ${\cal M}$ on extremal Kerr-Newman with $J$, 
with  $Q_{\mathrm{E}}$
and $Q_{\mathrm{M}}$ given. The details of this variational problem are 
discussed in \cite{GabJarRei11}, where rigidity is also proved.

\end{proof}

\section{Discussion}
\label{s:discussion}
We have reviewed a set of geometric inequalities holding for 
stably outermost marginally trapped surfaces embedded in
generic dynamical, non-necessarily axisymmetric spacetimes with
ordinary matter that can extend and cross the black hole horizon.
These inequalities provide lower bounds for the area $A$, in terms 
of expressions involving (linearly) the angular momentum $J$  and (quadratically)
the electric and magnetic charges, $Q_{\mathrm{E}}$ and $Q_{\mathrm{M}}$. 
Extensions including Yang-Mills charges, $Q^{\mathrm{YM}}_{\mathrm{E}}$ and $Q^{\mathrm{YM}}_{\mathrm{M}}$, 
as well as a charge $Q[\eta]$
for certain divergence-free vectors, have also been discussed. 
If $J$ is present, axisymmetry is required on the
surface (and only on the surface). Otherwise the inequalities involve no
symmetry requirements.

We have adopted a purely quasi-local spacetime Lorentzian approach.
However, it is worthwhile to note that these inequalities were initially
discussed on initial data in spatial 3-slices by using Riemannian techniques, 
in particular minimal surfaces. Although more stringent in their spacetime 
requirements, whenever applicable, such versions also 
hold on more general surfaces that marginally outer trapped surfaces.
We have however focused here on the specific context of black hole horizons.
In this setting, the adoption of a spacetime perspective based entirely on 
purely Lorentzian concepts has offered crucial geometric insights
into the problem: all geometric elements in the proof
acquire a clear spacetime meaning. This has lead to a refinement 
in the required conditions permitting, in particular, the generic incorporation 
of matter in the discussion. 
The crucial ingredient enabling the shift to a purely Lorentzian discussion 
has been the identification of the stably outermost condition 
for marginally outer trapped surfaces as the elementary involved notion.
In essence, this is the only required ingredient.
In this sense, the fulfillment of inequalities (\ref{e:inequality})
is just a fundamental and direct (irreducible) consequence of the 
Lorentzian structure of spacetime. This is the main conclusion that we want to stress in these
notes.

Strictly speaking, the inclusion of the angular momentum in the inequalities 
requires two further (related) elements: axisymmetry on the surface and an
analytical variational principle.
This is in contrast with inequalities in which  $J$ is absent, that are
straightforward geometric consequences of the stability 
condition. 
Certainly, the identification of potentials 
$\sigma$, $\omega$, $\chi$ and $\psi$ for the functional ${\cal M}$
is related to the spherical topology of ${\cal S}$, ultimately
controlled by the stability condition for MOTS.
However, it is indeed of relevance to assess the role of the axisymmetry 
and variational treatment requirements in this problem.
First, relaxing the local axisymmetry in the angular momentum 
characterization is of interest in astrophysical contexts. 
Second, the success of the variational problem is intrinsically 
related to the existence of particular spacetimes,
namely extremal stationary (axisymmetric) black holes, that 
saturate the inequality and simultaneously provide a (unique) minimum 
for the functionals ${\cal M}$. 
A better understanding of the structural role played by the
variational principle in the proofs could offer insight 
into the properties of the space of solutions of the theory.
In particular, the observation that spacetimes admitting symmetries are 
singular points in the space of solution of (vacuum) Einstein equations
\cite{Arms:1982ea,FisMarMon80} (namely conical singularities) 
could shed some light on the relation between the presence of symmetries
and the need of a variational principle.

From a physical perspective, stable marginally trapped surfaces are sections 
of quasi-local models for black hole horizons. More precisely, the 
spacetime stably outermost condition is essentially the {\em outer condition}
 introduced in \cite{Hay94} for trapping horizons, namely worldtubes of 
apparent horizons. From an initial data perspective, the (strictly) stably outermost condition
is precisely the condition that guarantees the evolution of an initial apparent
horizon into a dynamical horizon \cite{AndMarSim05,AndMarSim08} with a unique foliation by
 marginally outer trapped surfaces \cite{AshGal05}.
The inequalities here studied provide a characterization of the notion
of black hole horizon (sub)extremality \cite{Booth:2007wu}. Moreover, the
rigidity results imply that the saturation of the inequalities characterize the
extremality of the horizon geometry. These considerations 
endorse the discussion of the first law of thermodynamics in dynamical horizons
\cite{AshKri02,AshKri03} where, in particular, the positivity of the surface gravity
is equivalent to the fulfillment of the inequalities here discussed. Equivalently, 
support is given for the physical validity of the Christodoulou mass, 
as a function growing with the area (for fixed $J$ and $Q_i$). 
Beyond the inequalities among $A$, $J$ and the charges $Q_i$, but still in the 
context of energy balance equations,
we have noted in section \ref{s:AQflux} that the integral characterization 
of the stability condition can be interpreted as an {\em energy flux} inequality.

In the general context of the standard picture of gravitational collapse \cite{Pen73}, 
the inequalities here studied provide a set of quasi-local geometric probes into 
black hole dynamics in generic situations. In this sense, it is of interest to
explore a possible connection between these inequalities and aspects of the cosmic 
censorship conjecture
(e.g. through their link to related global inequalities \cite{GabJarRei11}),
or possible implications in the understanding of partial problems in black hole stability.

We would like to conclude by emphasizing that these inequalities
represent a particular example of the extension
to a Lorentzian setting of tools and concepts employed in the discussion of 
minimal surfaces in a Riemannian context. In this sense, this family of problems 
provides a concrete bridge between research in Riemaniann and Lorentzian
geometries.

\begin{acknowledgement}
This work is fully indebted to the close scientific collaboration with
S. Dain, M.E. Gabach Cl\'ement, M. Reiris and W. Simon. I would like express 
here my gratitude to them.
I would also like to thank  A. Ace\~na, M. Ansorg, C. Barcel\'o, M. Mars and J.M.M. Senovilla
for useful discussions. 
I thank  M.E. Gabach Cl\'ement for her careful reading of the manuscript.
I acknowledge the support of the Spanish MICINN 
(FIS2008-06078-C03-01) and the Junta de Andaluc\'\i a (FQM2288/219).

\end{acknowledgement}
%

%
%
%

\end{document}